\documentstyle[twoside,fleqn,espcrc2]{article}
\newcommand{\AmS}{{\protect\the\textfont2
  A\kern-.1667em\lower.5ex\hbox{M}\kern-.125emS}}

\title{
Low-Temperature Series for Ising Model by Finite-Lattice Method
       }

\author{
H. Arisue
    \address{
        Osaka Prefectural College of Technology,
        Saiwai-cho, Neyagawa, Osaka 572, Japan
             }%
 \ \ and \ K. Tabata
    \address{
        Osaka Institute of Technology,
        Junior College,
        Ohmiya, Asahi-ku, Osaka 535, Japan
             }%
        }

\begin{document}

\pagestyle{empty}

\begin{abstract}
We have calculated the low-temperature series
for the second moment of the correlation function
in $d=3$ Ising model to order $u^{26}$
and for the free energy of Absolute Value Solid-on-Solid (ASOS) model
to order $u^{23}$, using the finite-lattice method.
\end{abstract}

% typeset front matter (including abstract)
\maketitle

%%%%%%%%%%%%%%%%%%%%%%%%%%%%%%%%%%%%%%%%%%%%%%%%%%%%%%%%%%%%%%%
\section{INTRODUCTION }

 Recently the low-temperature series of $d=3$ Ising model
or equivalently strong-coupling series of $d=3$ $Z_2$
lattice gauge theory have been extended to higher orders
using finite-lattice method.
 The finite-lattice method to obtain series expansion
was originally developed by Neef and Enting
     \cite{Enting}.
 In this method the expansion series of the free energy density
in the infinite volume limit to an order in the expansion-parameter
is given by the appropriate linear combination of
the free energies on finite-size lattices.
 The coefficients of the linear combination are given
by M\"obius inversion
     \cite{Domb}.
 This procedure avoids the problem involved in the graphical method,
in which it is rather difficult
to give the algorithm for listing all the diagrams completely
that contribute to the relevant order of the series.
 The finite-lattice method is more effective in lower dimensions
     \cite{GuttmannPRL}
and it was applied intensively to two-dimensional systems
     \cite{Entingfour}.

 The M\"obius inversion was introduced to the field of
lattice gauge theory by Mack
     \cite{Mack}
and was used to make a partial resummation of the strong-coupling
expansion of the theory
      \cite{Mack,Muenster}.
 Fujiwara and one of the author ( H. A.) developed the finite-lattice
method of strong-coupling expansion and its full resummation
in lattice gauge theory using the M\"obius inversion
    \cite{Arisueone,Arisuetwo}.
The full-resummation method was applied to the calculation of
the free energy and string tension for $SU(2)$
    \cite{Arisueone}
and $Z_2$
    \cite{Arisuetwo}
lattice gauge theory in three dimensions,
and of the free energy for $Z_2$
    \cite{Hirata}
and $SU(2)$
    \cite{Narayanan}
lattice gauge theory in four dimensions.
 The method was also applied to obtain the strong-coupling expansion
series of the free energy to order $u^{20}$,
string tension to order $u^{13}$
    \cite{Arisuenine}
and the mass gap to order $u^{11}$
    \cite{Arisuethree}
in $d=3$ $Z_2$ lattice gauge theory,
and of the free energy in $d=4$ $Z_2$ lattice gauge theory
to order $u^{11}$
    \cite{Arisueeight},
where $u=\tanh^2{\beta_{gauge}}$
and $\beta_{gauge}$ is the inverse gauge coupling constant.
% As for the mass gap Decker had calculated the series to order $u^{}$

 The $d=3$ $Z_2$ lattice gauge theory is dual to the $d=3$ Ising model
and the strong-coupling series of the free energy, string tension
and mass gap in the former is exactly the same
as the low-temperature series of the free energy,
surface tension and true inverse correlation-length
in the latter, respectively,
if the expansion-parameter $u=\tanh^2{\beta_{gauge}}$ in the former
is read as $u=\exp{(-4\beta)}$ and $\beta=J/kT$ in the latter.
 Thus the finite-lattice method of strong-coupling expansion
in $d=3$ lattice gauge theory is exactly the same
as that of low-temperature expansion in $d=3$ Ising model.

 Recently the low-temperature series of $d=3$ Ising model
were extended to higher orders by the finite-lattice method
using transfer matrix formalism
for calculating the partition function
based on building up finite-size
lattices one site at a time, which was originally invented by Enting
     \cite{Entingtwo}.
The calculated quantities are
the free energy to order $u^{25}$
     \cite{Bhanot},
the free energy, magnetization and susceptibility to order $u^{26}$
     \cite{Guttmann},
the surface tension to order $u^{17}$
     \cite{Arisuefour}
and the true inverse correlation-length to order $u^{15}$
     \cite{Arisuefive}.

 We should mention that Vohwinkel
     \cite{Vohwinkel}
obtained low-temperature Ising series
for free energy, magnetization and susceptibility,
which are longer than those of the
reference~\cite{Bhanot} or \cite{Guttmann}
using a modification of
the shadow-lattice technique.
His method appears to be so powerful even in three dimensions,
although it is more efficient for higher dimensional systems.
We think, however, that
the finite-lattice method can still be the method
of choice for a range of problems in three dimensions
     \cite{GuttmannPRL}.

 We report here the application of the finite-lattice method
to calculate the low-temperature series
for the second moment of the correlation function
in $d=3$ Ising model to order $u^{26}$
      \cite{Arisueten}
and for the free energy of Absolute Value Solid-on-Solid (ASOS) model
to order $u^{23}$
      \cite{Arisueeleven}.
%%%%%%%%%%%%%%%%%%%%%%%%%%%%%%%%%%%%%%%%%%%%%%%%%%%%%%%%%%%%%%%
\section{
SECOND MOMENT
OF THE CORRELATION FUNCTION IN D=3 ISING MODEL}

 The point in the algorithm of the low-temperature expansion
for the second moment $\mu_2$ is the following.
 We consider the partition function
\begin{equation}
   Z(\beta,h,\eta,\gamma_1,\gamma_2,\gamma_3)
       = \sum_{ \{ s_i \} } \exp{( - {\cal H} )},    \label{eqn:Z}
\end{equation}
with the Hamiltonian
\begin{equation}
    {\cal H } = \beta \sum_{\langle ij \rangle} s_i s_j
            + \sum_i ( h + \gamma_1 x_i + \gamma_2 y_i + \gamma_3 z_i
                          + \eta \mbox{\boldmath$r$}_i^2 )  s_i,
\end{equation}
for the three-dimensional lattice with a volume $ V $.
 The second moment  is given by the second derivative of
the free energy density in the infinite-volume limit as
%\begin{eqnarray}
%\lefteqn{
%    \mu_2 =
%       \lim_{V \rightarrow \infty}     \frac{2}{V}
%    \left( \frac{\partial^2}{\partial h \partial \eta }
%        -  \frac{\partial^2}{\partial \gamma_1^2 }
%        -  \frac{\partial^2}{\partial \gamma_2^2 }
%        -  \frac{\partial^2}{\partial \gamma_3^2 }  \right)
%        }                                      \nonumber \\
%\lefteqn{
%    \times \ln{ Z(\beta,h,\eta,\gamma_1,\gamma_2,\gamma_3) }
%           |_{h=\eta=\gamma_1=\gamma_2=\gamma_3=0}.
%        }
%\end{eqnarray}
\[
    \mu_2 =
       \lim_{V \rightarrow \infty}     \frac{2}{V}
    \left( \frac{\partial^2}{\partial h \partial \eta }
        -  \frac{\partial^2}{\partial \gamma_1^2 }
        -  \frac{\partial^2}{\partial \gamma_2^2 }
        -  \frac{\partial^2}{\partial \gamma_3^2 }  \right)
\]
\vspace{-0.55cm}
\begin{equation}
%\lefteqn{
    \times \ln{ Z(\beta,h,\eta,\gamma_1,\gamma_2,\gamma_3) }
           |_{h=\eta=\gamma_1=\gamma_2=\gamma_3=0}.
%        }
\end{equation}
\vspace{-0.2cm}
 Then the finite-lattice method can be applied to the low-temperature
expansion of the free energy density, which should be calculated to
the order of $u^N h \eta$ or $u^N \gamma_i^2 (i=1,2,3)$
to obtain the second-moment series to $u^N$.

  We have obtained the series for the second moment
to order $u^{26}$,
extending the previous result of order $u^{15}$ calculated
by Tarko and Fisher
      \cite{Tarko}
using the standard graphical method
and of order $u^{19}$ calculated by Vohwinkel and Weisz
      \cite{Vohwinkeltwo}
using the shadow-lattice technique.
Vohwinkel and Weisz also gave an estimate of
the series to order $u^{29}$.
 Our exact series to order $u^{19}$ coincides with their exact result
and our exact coefficients from order $u^{20}$ to $u^{26}$ are
consistent with their estimate within an accuracy of 1 per cent
for each of the orders.
 It gives the low-temperature series
for the second-moment correlation length squared $\Lambda_2={\xi_1}^2$
to order $u^{23}$,
when combined with the known low-temperature series
of the susceptibility
      \cite{Guttmann,Vohwinkel}.
 This is longer by six terms
than the low-temperature series for the true correlation length
squared $\Lambda_2^{\prime}$
that was derived from the true inverse correlation-length
given in Ref.~\cite{Arisuefive}.
  An analysis of the obtained series
by inhomogeneous differential approximants
gives critical exponents
$ 2\nu^{\prime} + \gamma^{\prime} = 2.509(38)$
for the second moment
and
$ 2\nu^{\prime} = 1.247(19) $
for the correlation length squared.
 These are consistent with
the results from high-temperature series, $\epsilon$-expansion and
Monte Carlo analysis and we can conclude that the scaling relation
between the high- and  low-temperature exponents as
$\nu=\nu^{\prime}$ and $\gamma=\gamma^{\prime}$ is satisfied
within an accuracy of about 2 per cent. The details can be seen
in the reference~\cite{Arisueten}.

%%%%%%%%%%%%%%%%%%%%%%%%%%%%%%%%%%%%%%%%%%%%%%%%%%%%%%%%%%%%%%%
\section{
FREE ENERGY OF ASOS MODEL}

 In the low-temperature expansion of the surface tension
to order $u^{17}$
     \cite{Arisuefour}
we found that the series coefficients change their sign at
the order of $u^{13}$.
The surface tension in three-dimensional Ising model
or the string tension in three- or higher-dimensional
lattice gauge theory
suffers from the roughening transition
     \cite{Weekstwo,Hasenfratz,Itzykson}
and it is expected to exhibit Kosterlitz-Thouless type singularity like
\begin{equation}
   f(u)=A(u) \exp{[-c(u_r-u)^{-1/2}]} + B(u)
            \label{KT}.
\end{equation}
It has the essential singularity
at the roughening transition point $u_r$.
As was pointed out by Hasenbusch and Pinn
     \cite{Hasenbusch},
the sign-change is just the signal of the K-T type singularity
in equation~(\ref{KT}).
In fact if we expand the function~(\ref{KT}) in terms of $u$
we would obtain a series with a sign-change.
The order of the sign-change depends on $c$ and $u_r$.

 Absolute Value Solid-on-Solid (ASOS)  model is an approximation of
the interface of $d=3$ Ising model.
It neglects overhangs and disconnected parts and
is also expected to exhibit K-T type phase transition.
The free energy of ASOS model just corresponds to the surface tension
of $d=3$ Ising model and is expected to behave like~(\ref{KT}).
 Hasenbusch and Pinn
     \cite{Hasenbuschtwo}
calculated the low-temperature series
for the free energy of ASOS model to order $u^{12}$
using the finite-lattice method,
extending the previous work by Weeks et al
     \cite{Weeks}
and found the expected sign-change at the order of $u^{11}$.

 We have extended the low-temperature series for the free energy
to $u^{23}$ using the finite-lattice method
     \cite{Arisueeleven},
which is longer by 11 terms than that by Hasenbusch and Pinn.
%  <technique of expansion>
%
%
%
 In the longer series the sign-change is seen
only at the order of $u^{11}$ found by Hasenbusch and Pinn.
 We have fitted the obtained series to the Taylor expansion of
the fuction~(\ref{KT}) with $A(u)=constant$ and $B(u)=0$.
A good fitting is obtained
if we take $u_r=0.214$ and $ c = 0.527 $.
 This fitted value of $u_r$ should be compared with
$u_r=0.207(9)$ from the series analysis of the surface width
and
$u_r=0.1994(1)$ from the Monte Carlo Renormalization group analysis
      \cite{Hasenbuschthree}.
These results confirm that the roughening phase-transition
of ASOS model is of K-T type.

\end{document}